\input{aipcheck}

\documentclass[
    ,final         % use final for the camera ready runs
  ]
  {aipproc}

\layoutstyle{8x11single}

\begin{document}

\title{How large is the gluon polarization in the statistical parton distributions approach?}

\classification{12.40.Ee, 13.60.Hb, 13.88.+e, 14.70.Dj}
\keywords      {Gluon polarization; Proton spin; Statistical distributions}

\author{Jacques Soffer}{
  address={Physics Department, Temple University, 1835 N, 12 Street, Philadelphia, PA 19122-6082, USA}
}

\author{Claude Bourrely}{
  address={Aix-Marseille Universit\'e, D\'epartement de Physique, Facult\'e des Sciences, site de Luminy, 13288 Marseille, Cedex 09, France}
}

\author{Franco Buccella}{
  address={INFN, Sezine di Napoli, Via Cintia, Napoli, I-80126, Italy }
}

\begin{abstract}
We review the theoretical foundations of the quantum statistical approach to parton distributions and we show that by using some recent experimental results from Deep Inelastic Scattering, we are able to improve the description of the data by means of a new determination of the parton distributions. We will see that a large gluon polarization emerges, giving a significant contribution to the proton spin.
                   
\end{abstract}

\maketitle

%%%%%%%%%%%%%%%%%%%%%%%%%%%%%%%%%%%%%%%%%%%%
%% MAINMATTER
%%%%%%%%%%%%%%%%%%%%%%%%%%%%%%%%%%%%%%%%%%%%

\section{Introduction}
 Several years ago a new set of parton distribution functions (PDF) was constructed in the
framework of a statistical approach of the nucleon \cite{bbs1}. For quarks
(antiquarks), the building blocks are the helicity dependent distributions
$q^{\pm}(x)$ ($\bar q^{\pm}(x)$). This allows to describe simultaneously the
unpolarized distributions $q(x)= q^{+}(x)+q^{-}(x)$ and the helicity
distributions $\Delta q(x) = q^{+}(x)-q^{-}(x)$ (similarly for antiquarks). At
the initial energy scale, these distributions
are given by the sum of two terms, a quasi Fermi-Dirac function and a helicity
independent diffractive
contribution. The flavor asymmetry for the light sea, {\it i.e.} $\bar d (x) > \bar
u (x)$, observed in the data, is built in. This is simply understood in terms
of the Pauli exclusion principle, based on the fact that the proton contains
two up-quarks and only one down-quark. The chiral properties of QCD lead to
strong relations between $q(x)$ and $\bar q (x)$.
For example, it is found that the well established result $\Delta u (x)>0 $\
implies $\Delta
\bar u (x)>0$ and similarly $\Delta d (x)<0$ leads to $\Delta \bar d (x)<0$. 
This earlier prediction was confirmed by recent data. In addition we found the approximate equality of the flavor asymmetries, namely $\bar d(x) - \bar u(x) \sim \Delta \bar u(x) - \Delta \bar d(x)$. \\
After a short review of the statistical approach, we will give some results on polarized Deep Inelastic Scattering (DIS) and the gluon helicity distribution, before
turning to our predictions for helicity asymmetries recently measured at RHIC-BNL with the polarized $pp$ collider.

\section{Basic review on the quantum statistical approach}
Let us first recall some of the basic components for building up the parton
distribution functions (PDF) in the statistical approach, as oppose to the
standard polynomial type parametrizations, based on Regge theory at low $x$ and counting rules at large $x$. The fermion distributions are given by the sum of two terms \cite{bbs1}, the
first one, 
a quasi Fermi-Dirac function and the second one, a flavor and helicity
independent diffractive
contribution equal for light quarks. So we have, at the input energy scale
$Q_0^2$,
\begin{equation}
xq^h(x,Q^2_0)=
\frac{AX^h_{0q}x^b}{\exp [(x-X^h_{0q})/\bar{x}]+1}+
\frac{\tilde{A}x^{\tilde{b}}}{\exp(x/\bar{x})+1}~,
\label{eq1}
\end{equation}
\begin{equation}
x\bar{q}^h(x,Q^2_0)=
\frac{{\bar A}(X^{-h}_{0q})^{-1}x^{\bar b}}{\exp [(x+X^{-h}_{0q})/\bar{x}]+1}+
\frac{\tilde{A}x^{\tilde{b}}}{\exp(x/\bar{x})+1}~.
\label{eq2}
\end{equation}
It is important to remark that $x$ is indeed the natural variable, and not the energy like in statistical mechanics, since all sum rules we will use are expressed in terms of $x$.
Notice the change of sign of the potentials
and helicity for the antiquarks.
The parameter $\bar{x}$ plays the role of a {\it universal temperature}
and $X^{\pm}_{0q}$ are the two {\it thermodynamical potentials} of the quark
$q$, with helicity $h=\pm$. We would like to stress that the diffractive
contribution 
occurs only in the unpolarized distributions $q(x)= q_{+}(x)+q_{-}(x)$ and it
is absent in the valence $q_{v}(x)= q(x) - \bar {q}(x)$ and in the helicity
distributions $\Delta q(x) = q_{+}(x)-q_{-}(x)$ (similarly for antiquarks).
The {\it nine} free parameters \footnote{$A$ and $\bar{A}$ are fixed by the following normalization conditions $u-\bar{u}=2$, $d-\bar{d}=1$.} to describe the light quark sector ($u$ and $d$), namely $X_{u}^{\pm}$, $X_{d}^{\pm}$, $b$, $\bar b$, $\tilde b$, $\tilde A$ and $\bar x$ 
in the above expressions, were
determined at the input scale from the comparison with a selected set of
very precise unpolarized and polarized DIS data 
\cite{bbs1}. The additional factors $X_{q}^{\pm}$ and $(X_{q}^{\pm})^{-1}$ come from the transverse momentum dependence (TMD), as explained in Refs.~[2,3]. For the gluons we consider the black-body inspired expression
\begin{equation}
xG(x,Q^2_0)=
\frac{A_Gx^{b_G}}{\exp(x/\bar{x})-1}~,
\label{eq5}
\end{equation}
a quasi Bose-Einstein function, with $b_G$, the only free parameter, since $A_G$ is determined
by the momentum sum rule. In our early works we were assuming for the polarized gluon 
distribution $x\Delta G(x,Q^2_0)= 0$, but in our recent analysis of a much larger set of data we had to give up this simplifying assumption, as we will see below. For the strange quark
distributions, the simple choice made in Ref.~[1]
was greatly improved in Ref.~[4]. Our procedure allows to construct simultaneously the unpolarized quark distributions and the helicity distributions. This is worth noting because it is a very unique situation. Following our first
paper in 2002, new tests against experimental (unpolarized and
polarized) data turned out to be very satisfactory, in particular in hadronic
collisions, as reported in Refs.~[5,6].

\section{Deep inelastic scattering}

\begin{figure}[hb]
 \includegraphics[height=.24\textheight]{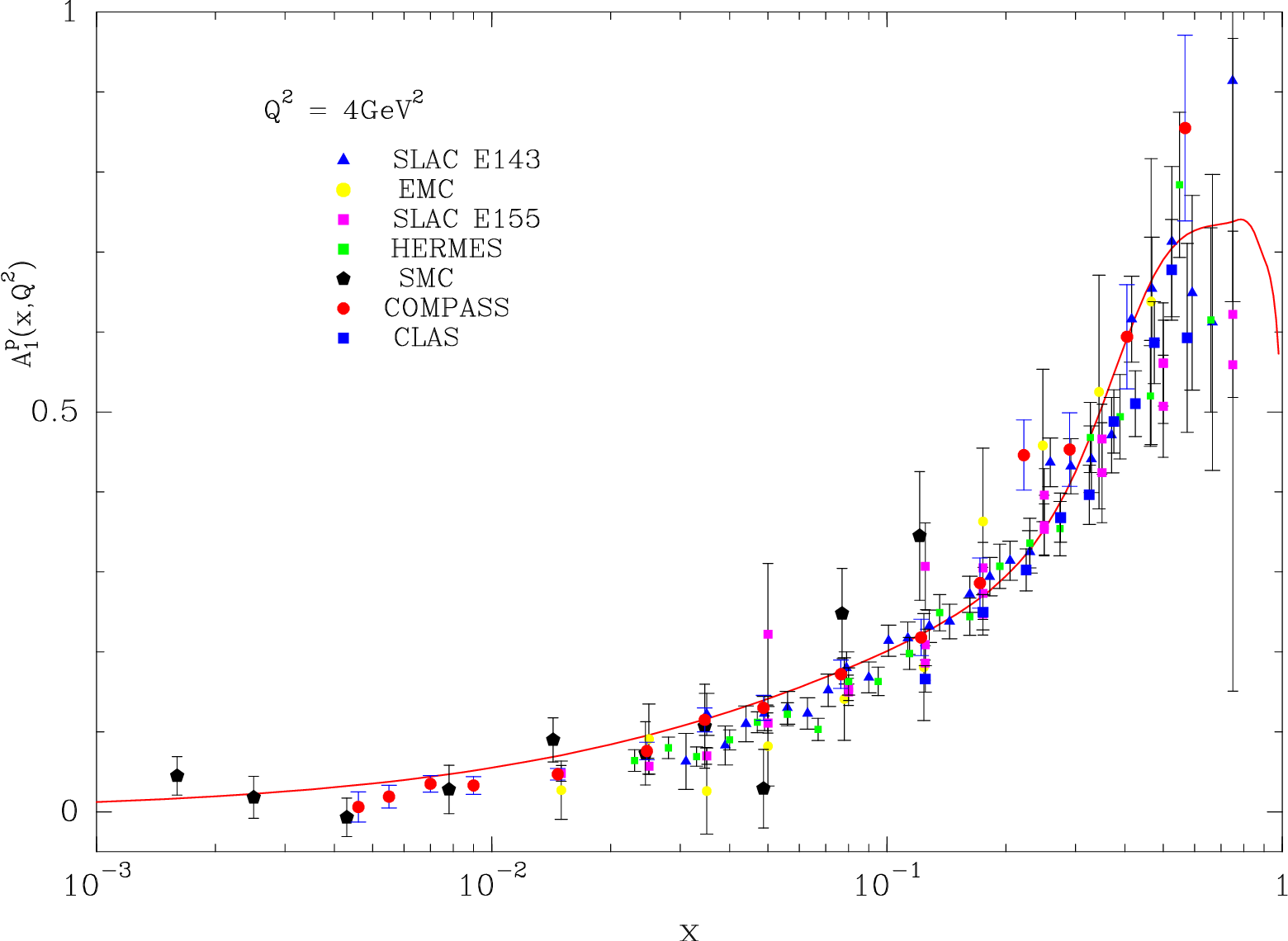}
  \includegraphics[height=.24\textheight]{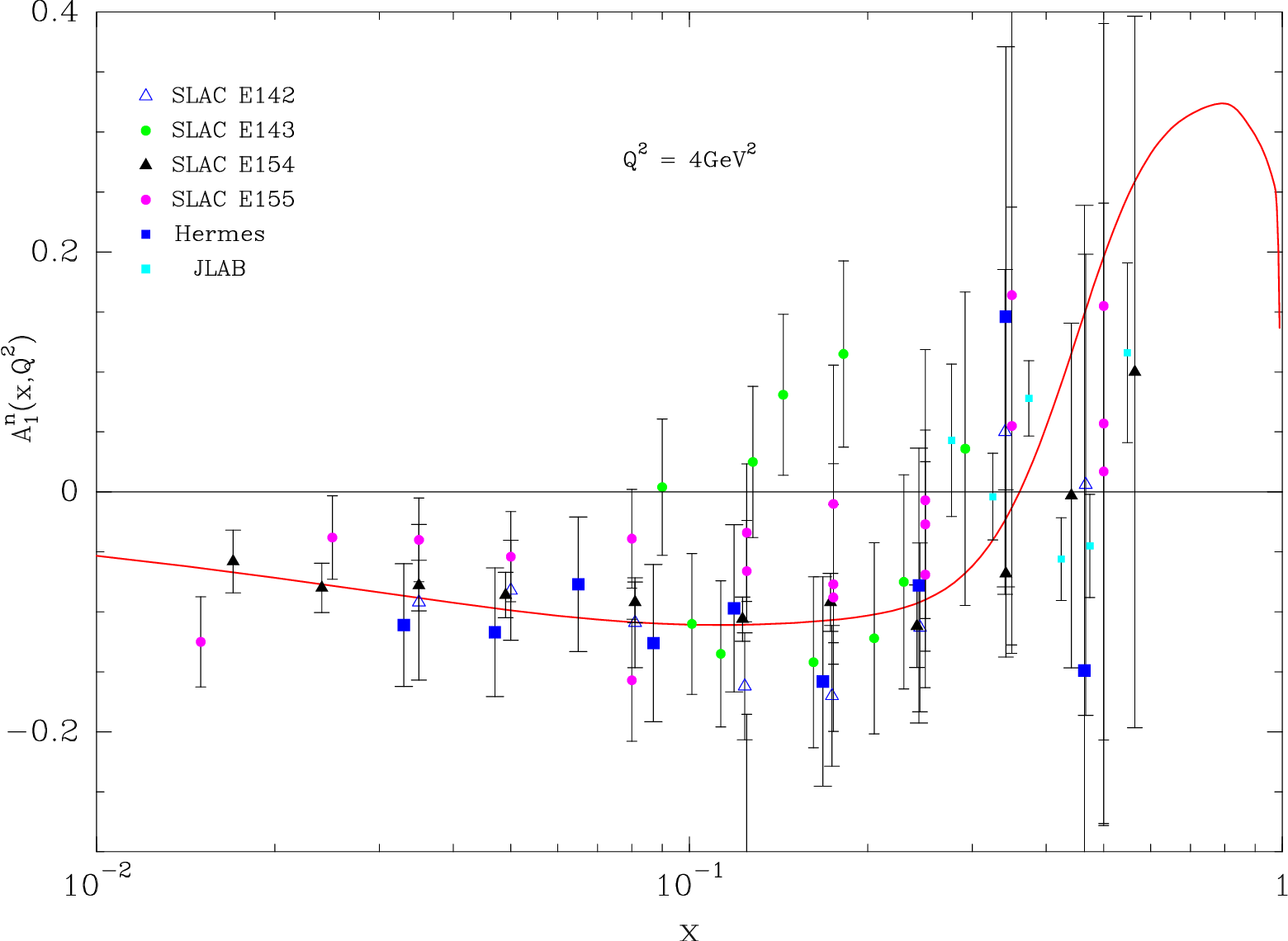}
  \caption{{\it  Left}: Comparison of the world data on $A_1^p(x,Q^2)$ at $Q^2= 4~\mbox{GeV}^2$, with the result of the statistical approach.
{\it Right}: Comparison of the world data on $A_1^n(x,Q^2)$ at $Q^2= 4~\mbox{GeV}^2$, with the result of the statistical approach.}
\label{fig1}
\end{figure}

First we consider the important issue of
the asymmetries $A_1^{p,d,n}(x,Q^2)$, measured in polarized DIS.
We recall the definition of the asymmetry $A_1(x,Q^2)$, namely
\begin{equation}
A_1(x,Q^2)= \frac{[g_1(x,Q^2) - \gamma^2(x,Q^2)g_2(x,Q^2)]}{F_2(x,Q^2)}
\frac{2x[1+R(x,Q^2)]}{[1+\gamma^2(x,Q^2)]}~,
\label{26}
\end{equation}
where $g_{1,2}(x,Q^2)$ are the polarized structure functions, $\gamma^2(x,Q^2)=4M^2x^2/Q^2$ and $R(x,Q^2)$ is the ratio between the
longitudinal and transverse photoabsorption cross sections. We display in Fig.~\ref{fig1} the world data on $A_1^{p,n}(x,Q^2)$ at $Q^2= 4~\mbox{GeV}^2$, with the new results of the statistical approach. Note that these asymmetries do NOT reach 1 when $x \to 1$, as required by the counting rules prescription, which we do not impose.

These results were obtained from a next-to-leading (NLO) fit of a large set of accurate DIS data (the unpolarized structure functions $F_2^{p,n,d}(x,Q^2)$, the polarized structure functions $g_1^{p,n,d}(x,Q^2)$, the structure function $xF_3^{\nu N}(x,Q^2)$ in $\nu N$ DIS, etc...) a total of 2140 experimental points with an average $\chi^2/pt$ of 1.33. For the polarized structure functions, it is slightly better since for a total of 271 experimental points, we have an average $\chi^2/pt$ of 1.21. Although the full details of these new results will be presented in a forthcoming paper, we just want to make a general remark. By comparing with the results of 2002 \cite{bbs1}, we have observed a remarquable stability of some important parameters, the light quarks potentials $X_{0u}^{\pm}$ and  $X_{0d}^{\pm}$, whose numerical values are almost unchanged. The new temperature is slightly lower. As a result the main features of the new light quark and antiquark distributions are only hardly modified, which is not surprizing, since our 2002 PDF set has proven to have a rather good predictive power.\\
The major novelty of this new analysis is the discovery of a large gluon helicity distribution, which has the specificity to be positive for all $x$ values. We display in Fig.~2 ({\it Left}) the gluon helicity distribution versus $x$ at the initial scale $Q_{0}^2 = 1~\mbox{GeV}^2$ and $Q^2 = 10~\mbox{GeV}^2$. At the initial scale it is sharply peaked around $x=0.4$, this feature lessens after some evolution but it remains positive. It is interesting to mention that a large gluon helicity distribution has been also found in recent works \cite{dssv2,nocera}.\\
We display $\Delta G(x,Q^2)/G(x,Q^2)$ in  Fig.~2 ({\it Right}) for two $Q^2$ values and some data points suggesting that the gluon helicity distribution is positive indeed. According to the counting rules constraints, this ratio should go to one when $x=1$, but we observe that this is not the case here. In some other parameterizations in the current literature, this ratio goes to zero, since the large $x$ behavior of $x \Delta G(x)$ is $(1 - x)^{\beta}$ with $\beta \gg 3$. Clearly one needs a better knowledge of $\Delta G(x,Q^2)/G(x,Q^2)$ for $x > 0.2$.\\

\begin{figure}
 \includegraphics[height=.325\textheight]{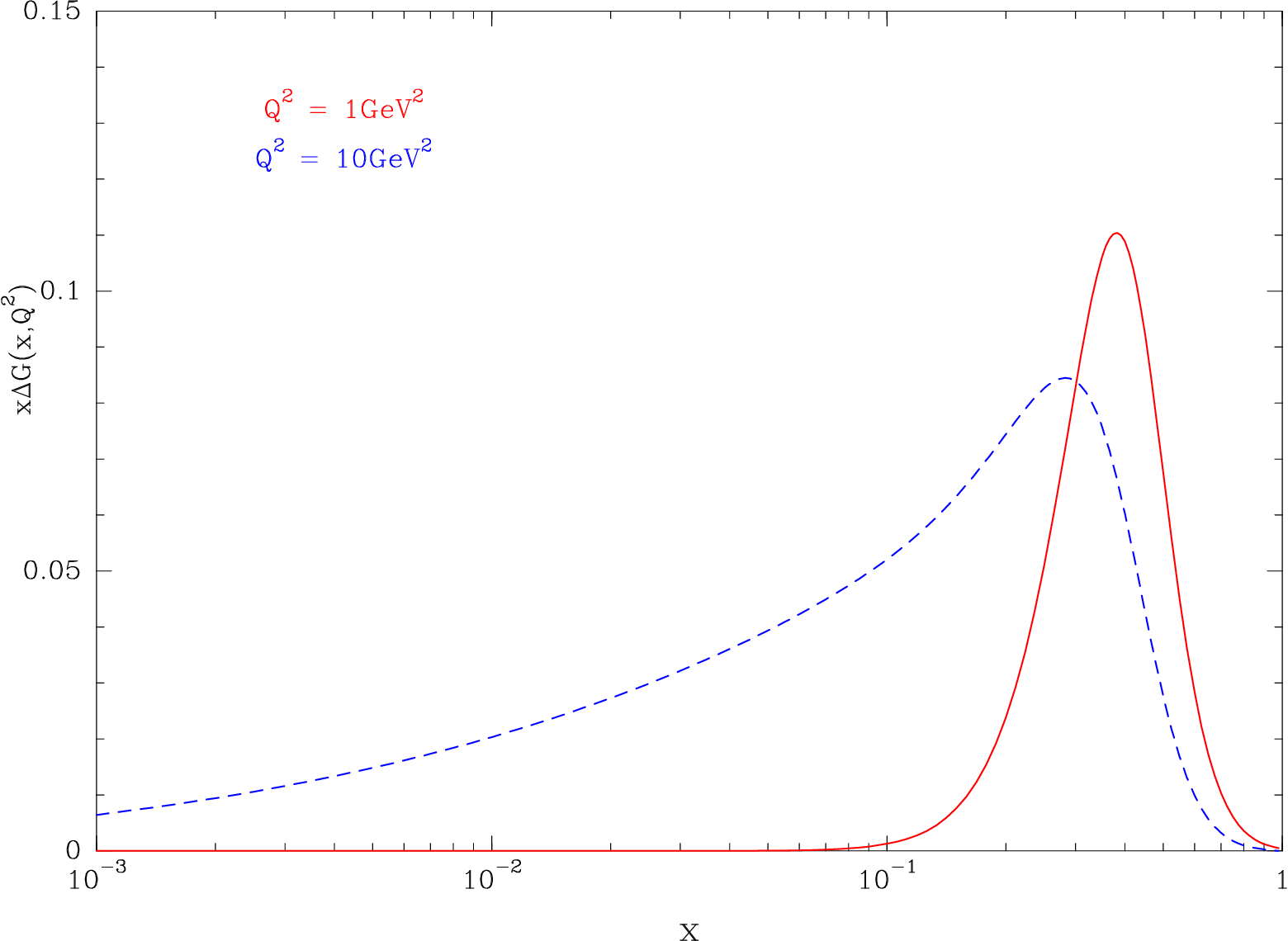}
  \includegraphics[height=.325\textheight]{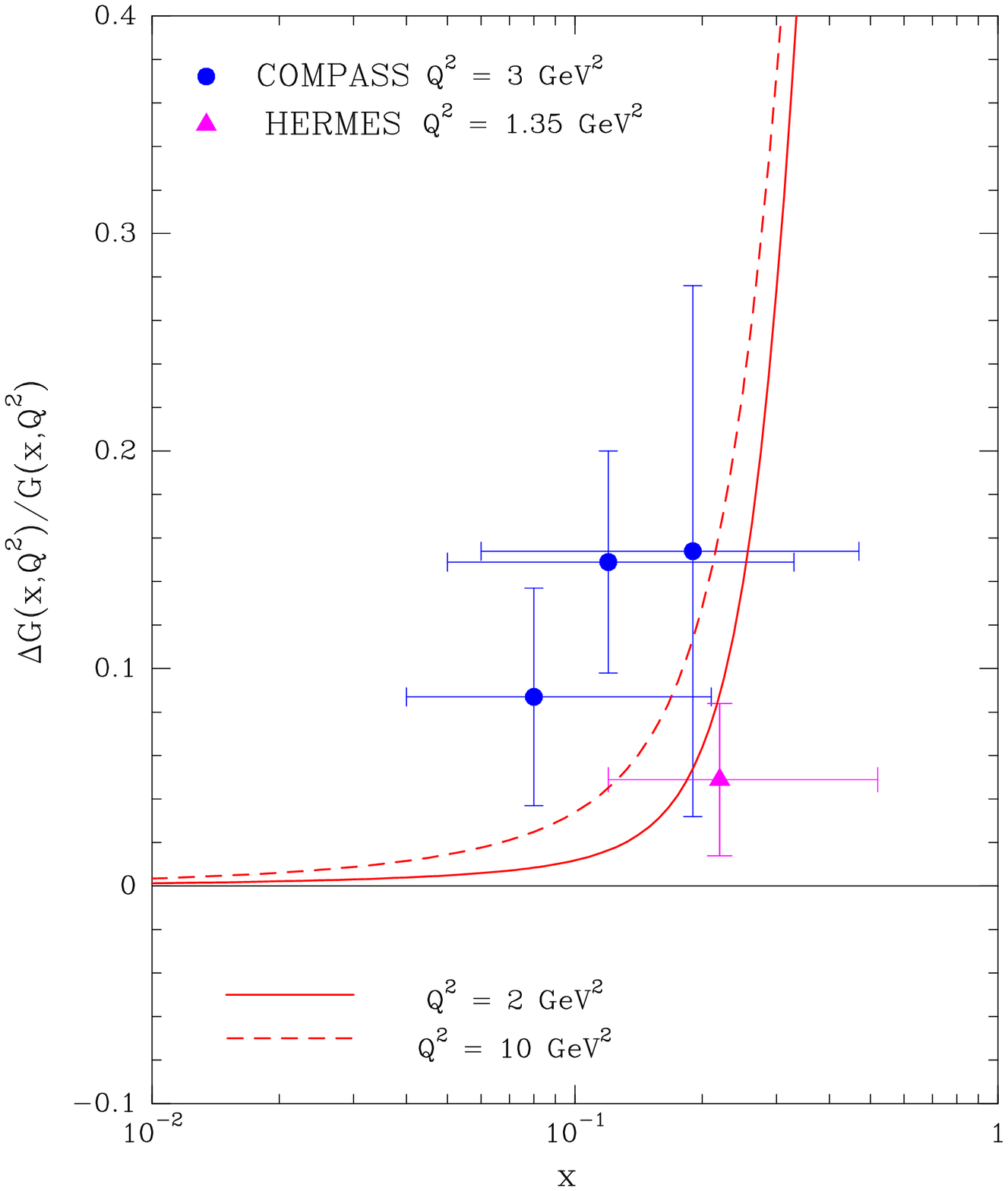}
  \caption{{\it Left}: The gluon helicity distribution $x\Delta G(x,Q^2)$ versus $x$, for $Q^2=1~\mbox{GeV}^2$ (dashed curve) and $Q^2=10~\mbox{GeV}^2$ (solid curve) (Taken from Ref.~\cite{bs14}). {\it Right}: $\Delta G(x,Q^2)/G(x,Q^2)$ versus $x$, for $Q^2=2~\mbox{GeV}^2$ (solid curve) and $Q^2=10~\mbox{GeV}^2$ (dashed curve) (Taken from Ref.~\cite{bs14}). }
	\label{fig2}
\end{figure}
%\newpage
\section{Helicity asymmetries at RHIC-BNL}
\begin{figure}[hb]
 \includegraphics[height=.25\textheight]{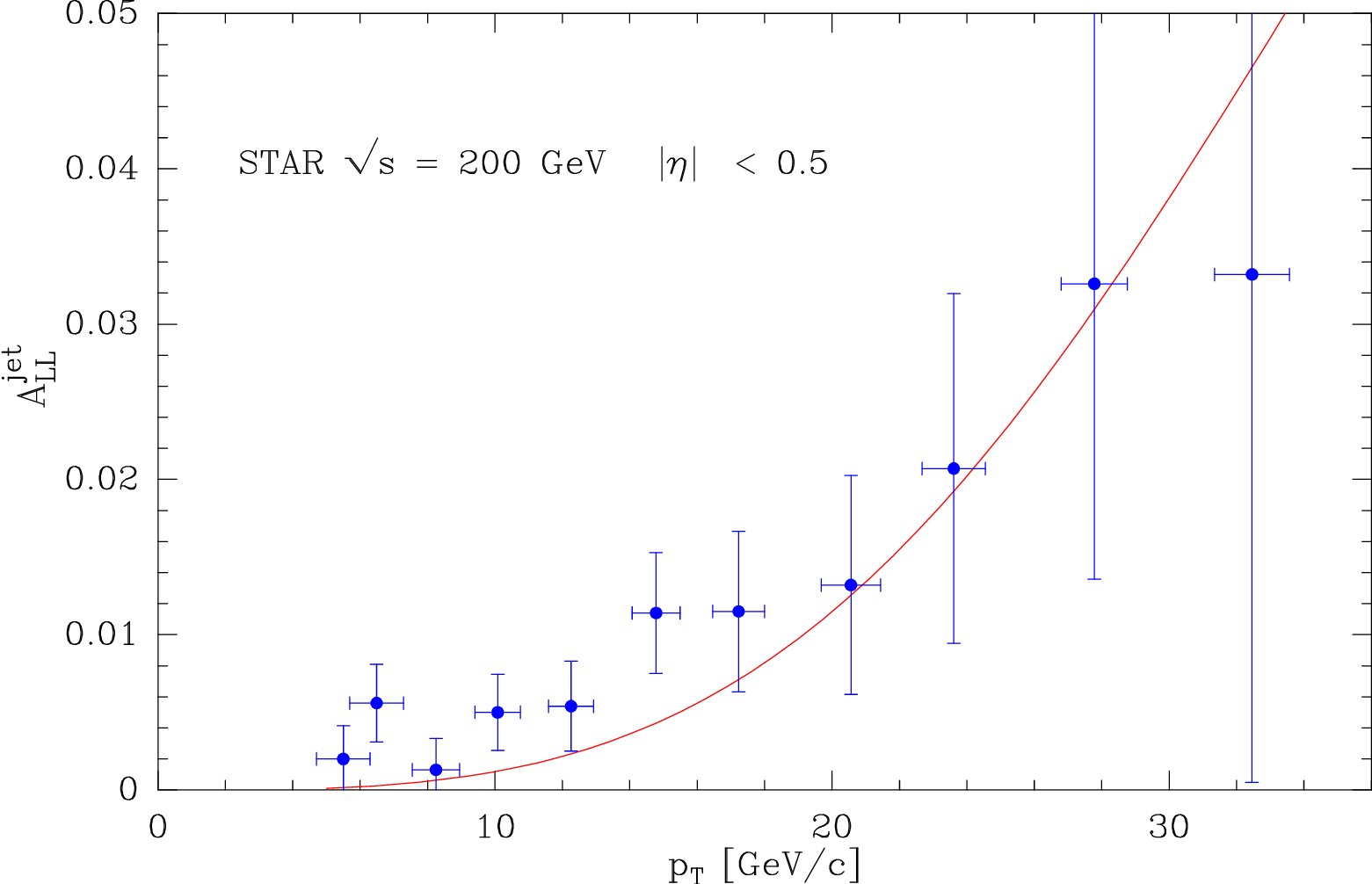}
  \includegraphics[height=.25\textheight]{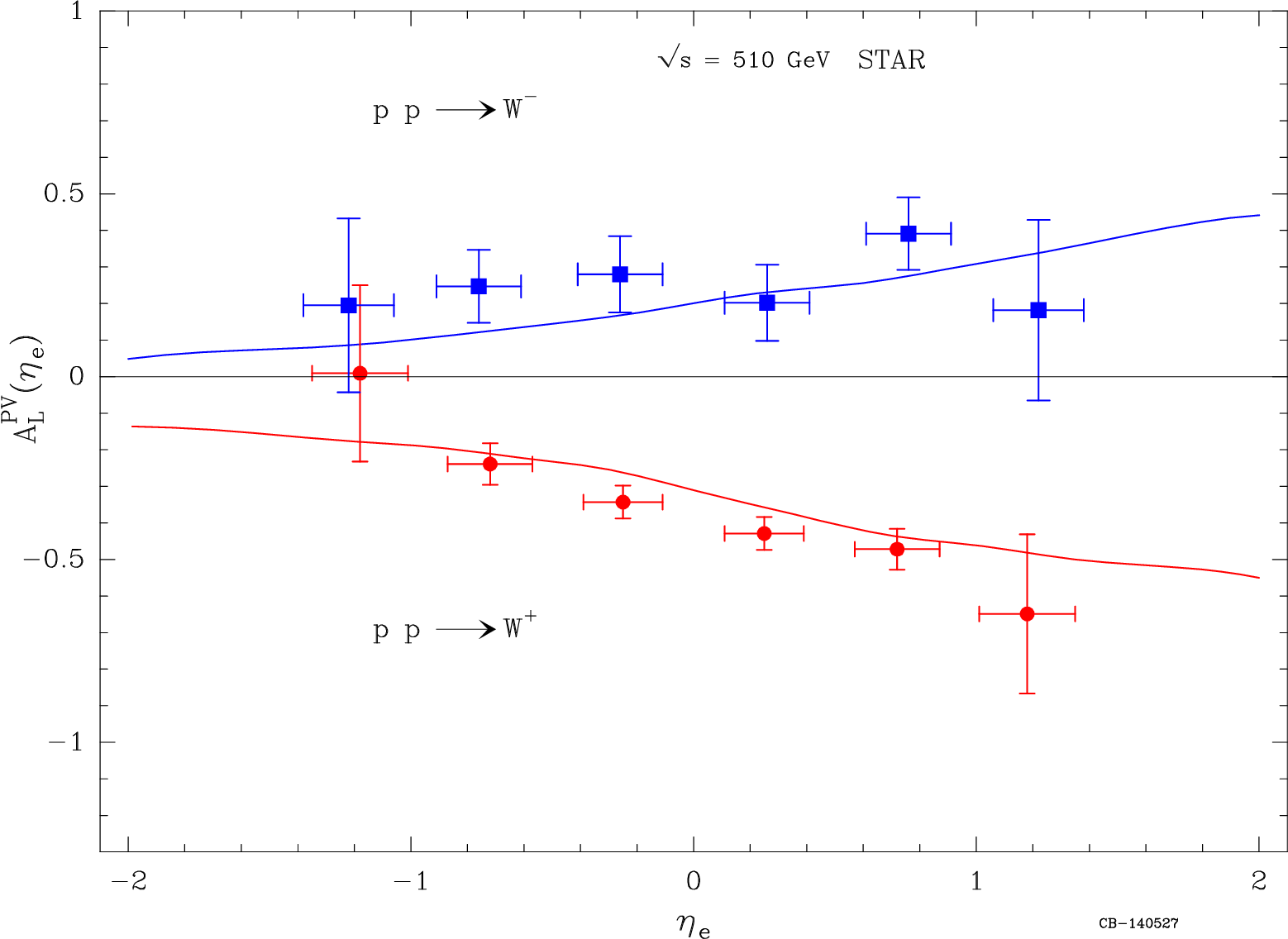}
  \caption{{\it Left}: Our predicted double-helicity asymmetry $A_{LL}^{jet}$ for jet production at BNL-RHIC in the near-forward rapidity region, versus $p_T$ (Taken from Ref. \cite{bs14}). {\it Right}: The measured parity-violating helicity asymmetries $A_L^{PV}$ for charged-lepton production at RHIC-BNL from STAR \cite{star2}, through production and decay of $W^{\pm}$ versus $y_e$, the charged-lepton rapidity. The solid curves are the predictions from the statistical approach.}
	\label{fig3}
	\end{figure}

Finally, we would like to test this new positive gluon helicity distribution in a pure hadronic reaction. In a very recent paper, the STAR Collaboration
at BNL-RHIC has reported the observation, in one-jet inclusive production, of a non-vanishing positive double-helicity asymmetry $A_{LL}^{\it jet}$ for $5 \leq p_T \leq 30 ~\mbox{GeV}$, in the near-forward rapidity region \cite{star1}. We display in Fig. \ref{fig3} ({\it Left}) our prediction  compared with these high-statistics data points and the agreement is very reasonable.\\ 
Next we consider the process $\overrightarrow p p\to W^{\pm} + X \to e^{\pm} + X$, where the arrow denotes a longitudinally polarized proton and the outgoing $e^{\pm}$ have been produced by the leptonic decay of the $W^{\pm}$ boson. The helicity asymmetry is defined as
\begin{equation}
A_L^{PV} = \frac{d\sigma_+ - d\sigma_-}{d\sigma_+  + d\sigma_-}~.
\label{AL}
\end{equation}
Here $\sigma_h$ denotes the cross section where the initial proton has helicity $h$. It was measured recently at RHIC-BNL \cite{star2} and the results are shown in Fig.~\ref{fig3} ({\it Right}). As explained in Ref. \cite{bbs13}, the $W^-$ asymmetry is very sensitive to the sign and magnitude of $\Delta \bar u$, so this is another successful result of the statistical approach.\\

%%%%%%%%%%%%%%%%%%%%%%%%%%%%%%%%%%%%%%%%%%%%
%% SAMPLE TABLE
%%
%% Shows the use of \tablehead and \tablenote
%% macros
%%%%%%%%%%%%%%%%%%%%%%%%%%%%%%%%%%%%%%%%%%%%

%%%%%%%%%%%%%%%%%%%%%%%%%%%%%%%%%%%%%%%%%%%%%%%%
%% BACKMATTER
%%%%%%%%%%%%%%%%%%%%%%%%%%%%%%%%%%%%%%%%%%%%%%%%


\bibliographystyle{aipproc}   % if natbib is available
\bibliographystyle{aipprocl} % if natbib is missing

%\bibliography{sample}

\begin{thebibliography}{9}


\bibitem{bbs1} C. Bourrely, F. Buccella and J. Soffer, \emph{Eur. Phys. J.} \textbf{C23}, 487 (2002).

\bibitem{bbs6} C. Bourrely, F. Buccella and J. Soffer, \emph{Phys. Rev.} \textbf{D83}, 074008 (2011).

\bibitem{bbs5} C. Bourrely, F. Buccella and J. Soffer, \emph{Int. J. of Mod. Phys.} \textbf{A28}, 1350026 (2013).

\bibitem{bbs2} C. Bourrely, F. Buccella and J. Soffer, \emph{Phys. Lett.}  \textbf{B648}, 39 (2007).

\bibitem{bbs3}  C. Bourrely, F. Buccella and J. Soffer, \emph{Mod. Phys. Lett.} \textbf{A18}, 771 (2003).

\bibitem{bbs4}  C. Bourrely, F. Buccella and J. Soffer, \emph{Eur. Phys. J.} \textbf{C41}, 327 (2005).

\bibitem{bs14}  C. Bourrely and J. Soffer, \emph{Phys. Lett.} \textbf{B740}, 168 (2015).

\bibitem{dssv2} D. de Florian, R. Sassot, M. Stratmann and W. Vogelsang, \emph{Phys. Rev. Lett.} \textbf{113}, 012001 (2014).

\bibitem{nocera} E. R. Nocera {\it et al.} (NNPDF Collaboration), arXiv:1406.5539v2 [hep-ph].

\bibitem{star1} STAR Collaboration, L. Adamczyk {\it et al.}, arXiv:1405.5134 [hep-ex].

\bibitem{star2} STAR Collaboration, L. Adamczyk {\it et al.}, \emph{ Phys. Rev. Lett.} \textbf {113}, 072301 (2014).

\bibitem{bbs13} C. Bourrely, F. Buccella and J. Soffer, \emph{Phys. Lett.}  \textbf{B726}, 296 (2013).




\end{thebibliography}

%{\typeout{}
  %\typeout{******************************************}
  %\typeout{** Please run "bibtex \jobname" to optain}
  %\typeout{** the bibliography and then re-run LaTeX}
  %\typeout{** twice to fix the references!}
  %\typeout{******************************************}
  %\typeout{}
% }

%\end{document}

%%%%%%%%%%%%%%%%%%%%%%%%%%%%%%%%%%%%%%%%%%%
%% The following lines show an example how to produce a bibliography
%% without the help of the BibTeX program. This could be used instead
%% of the above.
%%%%%%%%%%%%%%%%%%%%%%%%%%%%%%%%%%%%%%%%%%%

\end{document}